\documentclass[preprint,floats,aps,epsfig,nofootinbib,amssymb]{revtex4-1}
\usepackage{mathrsfs}
\usepackage{slashed}
\usepackage{graphicx,color}
\usepackage{dcolumn}
\usepackage{bm}
\usepackage{subfig}
\usepackage{graphicx}
\usepackage{amssymb}
\usepackage{stackrel}
\usepackage{pgfplots}
\usepackage{amsmath}
\usepackage{cases}
\usetikzlibrary{positioning}
\usetikzlibrary{snakes}
\usepackage{caption}

\newcommand\be{\begin{equation}}
\newcommand\ee{\end{equation}}
\newcommand\bea{\begin{eqnarray}}
\newcommand\eea{\end{eqnarray}}
\allowdisplaybreaks[4]

\begin{document}

\title{ Majorana Majoron and  the Baryon Asymmetry of the Universe}
\author{Wei Chao$^1$}
\email{chaowei@bnu.edu.cn}
\author{Ying-Quan Peng$^1$}
\email{yqpenghep@mail.bnu.edu.cn}
\affiliation{$^1$Center for Advanced Quantum Studies, Department of Physics, Beijing Normal University, Beijing, 100875, China}
\vspace{3cm}

\begin{abstract}

The spontaneous breaking of the global lepton number, an accidental symmetry in the Standard Model of particle physics, results in a massless goldstone boson, the Majoron, which can be taken as a cold dark matter candidate with properties similar to these of the axion. In this letter, we propose a novel mass generation mechanism for the Majoron  via radiative corrections induced by the interaction of a tiny lepton-number-violating (LNV) Majorana mass term of right-handed neutrinos in the canonical seesaw mechanism.  We show that this LNV Majorana mass term not only generates the mass of the Majoron but also leads to a non-zero initial velocity of the Majoron, which subsequently impacts on the relic abundance of the Majoron generated in the early universe via the misalignment mechanism. With the assistance of the Weinberg operator, the same initial velocity may also generate the lepton asymmetry, which is subsequently transported into the baryon asymmetry of the universe (BAU) via the weak sphaleron process. As a result, the neutrino masses, dark matter and the BAU can be addressed in this  concise theoretical framework.

\end{abstract}

\maketitle
\section{Introduction}

The discovery of the Higgs boson at the LHC~\cite{ATLAS:2012yve,CMS:2012qbp} is a milestone for the high energy physics. All particles predicted by the Standard Model (SM) of particle physics have been discovered. However,  the SM can only be a low-energy effective theory rather than a complete theory, considering that it can not explain all phenomena observed in particle physics experiments and astrophysical observations. 
New physics beyond the SM is needed to address at least the following three problems: the origin of  active neutrino masses~\cite{Formaggio:2021nfz}, dark matter (DM)~\cite{Bertone:2016nfn} and the baryon asymmetry of the Universe (BAU)~\cite{Bodeker:2020ghk}.  
It has been proven by facts that the symmetry is a wise tool for the model buildings of new physics. Considering that the lepton number (\textbf{L}), the baryon number (\textbf{B}) as well as \textbf{ B}$-$\textbf{ L} are accidental  global U(1) symmetries~\cite{Mohapatra:1980qe,Wetterich:1981bx,FileviezPerez:2010gw,Dulaney:2010dj,Chao:2010mp,Duerr:2013dza,Chao:2016avy,Chao:2015nsm} in SM, it makes sense to construct new physics theories based on these well-known symmetries that may account for as many new physics phenomena as possible, by assuming that new physics are minimal extensions to the SM according to the principle of the  Occam Razor.

In this letter, we propose an economical scenario that may address the origin of neutrino masses, DM and BAU problems simultaneously by taking a minimal extension to the canonical seesaw mechanism~\cite{Yanagida:1979as,GellMann:1979vob,Minkowski:1977sc,Mohapatra:1979ia}. Our theoretical framework is based on the following three conjectures:  (1) The $U(1)_{\textbf{L}}$ can be spontaneous broken by a non-zero vacuum expectation value (VEV) of a scalar singlet carrying nonzero \textbf{L}, which will also generate a Majorana mass term for right-handed neutrinos via the Yukawa interaction; 
(2) The tiny but nonzero active neutrino masses may origin from the  canonical seesaw mechanism. The decouple of heavy neutrinos leads to the dimension-5 Weinberg operator~\cite{Weinberg:1979sa}, which gives rise to neutrino Majorana masses after the breaking of the electroweak symmetry; and (3)  Global symmetries are expected to be explicitly broken by the quantum gravity effect~\cite{Alvey:2021hjp}, which means that a tiny lepton-number-violating (LNV) Majorana mass term for right-handed neutrinos, that is consistent with 't Hooft's naturalness criteria~\cite{tHooft:1979rat}, may naturally exist in a spontaneous breaking $U(1)_{\textbf{L}}$ theory.

According to the Goldstone theorem~\cite{Nambu:1961tp,Goldstone:1961eq,Goldstone:1962es}, there is a massless Goldstone boson, named Majoron, after the spontaneous breaking of the $U(1)_{\textbf{L}}$~\cite{Georgi:1981pg,Gelmini:1980re,Schechter:1981cv,Chikashige:1980ui}.  We show that the LNV Majorana mass term of right-handed neutrinos may lead to a tiny mass term for the Majoron via radiative corrections, which has never been noticed before and can be taken as a novel mass generation mechanism for axion-like particles (ALP). We dub the ALP with mass generated by this mechanism as the Majorana ALP. We  derive the effective potential of the Majoron explicitly and calculate its relic abundance by solving its equation of motion (EOM) numerically. In this case the Majoron is generated  via the Misalignment mechanism \cite{Dine:1982ah, Preskill:1982cy,Abbott:1982af,Co:2019jts,Chang:2019tvx,Batell:2021ofv,Chun:2021uwr,Chao:2022blc}.  Nevertheless,  the LNV term leads to an initial velocity for Majoron, similar to the case of the  kinetic misalignment mechanism~\cite{Co:2019jts,Chang:2019tvx}.   We show that the whole DM may be interpreted as Majoron and a large initial velocity of the Majoron may modify the Majoron oscillation time.

As another important by-product of the LNV Majorana mass term, a net  \textbf{L} carried by the Majoron can be produced  according to the Noether theorem.   Considering that the Majoron couples to leptons as well as $W\widetilde{W}$ and $B\widetilde{B}$ via triangle anomalies, the  \textbf{L} carried by Majoron may be subsequently transported into the $\textbf{B-L}$ asymmetry via the Weinberg operator and the  Chern-Simons operator. We show that the observed BAU  can be interpreted by the spontaneous baryogenesis mechanism \cite{Cohen:1987vi,Cohen:1988kt,Chiba:2003vp,Takahashi:2003db,Kusenko:2014uta,Ibe:2015nfa,Takahashi:2015waa,Jeong:2018jqe,Bae:2018mlv,Co:2019wyp,Domcke:2020kcp} with the Majoron as the source term, by solving the transport equations numerically.   As a result, the neutrino mass, DM and BAU can be simultaneously explained in the canonical seesaw mechanism plus Majoron picture.  It should be mentioned that our mechanism can be extended to the QCD axion scenario  in case that heavy fermions lie in the adjoint representation of the $SU(3)_{C}$ and its has a tiny mass term that explicitly violate the Peccei-Quinn symmetry. It provides a novel investigation direction for ALPs.

The remaining of the paper organized as follows: In section II we calculate the Majoron mass and its relic abundance in detail. Section III is devoted to the study of the BAU. The last part is concluding remarks.

\section{Majoron dark matter}
\label{sec:MajoronDM}

\begin{table}[t]
\begin{center}
\begin{tabular}{ c | c | c | c}
\hline
$a_1$ & $a_2$ & $a_3$ & $a_4$ \\ 
\hline
$m M^3 \left( 1+ \log {m^2 \over \mu^2 } \right)$ & $2 m^2 M^2 \log{m^2 \over \mu^2 }$ & $ m^3 M $ & $ m^4$ \\
\hline
\end{tabular}
\caption{Couplings in the Majoron potential, where $M$ represent heavy neutrino Majorana mass and $\mu$ is the energy scale. \label{tab:Couplings}}
\label{fr4}
\end{center}
\end{table}

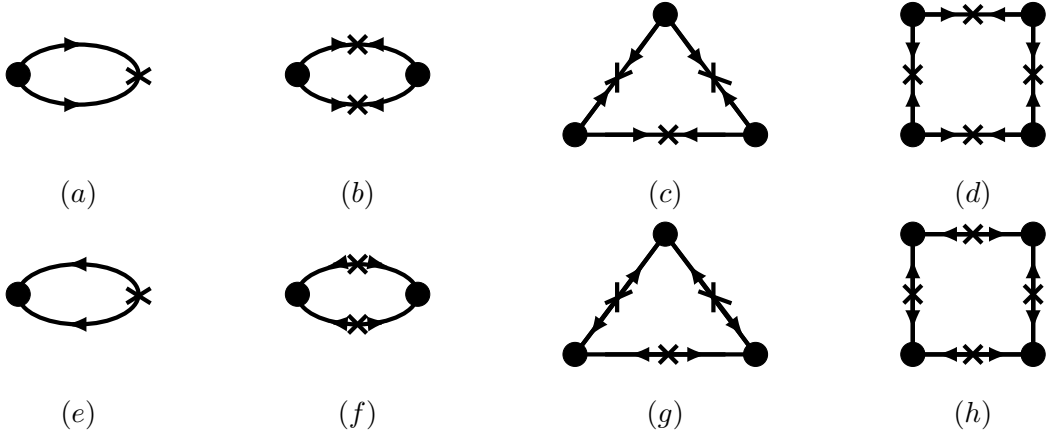
\begin{figure}
\begin{center}
\begin{tikzpicture}[scale=0.8]
\draw [-, ultra thick] (1.8,0.85) -- (2.2,1.1);
\draw [-, ultra thick] (1.8,1.1) -- (2.2,0.85);
\draw [fill](0,1) circle (0.2);
\draw[style,ultra thick] (1,1) ellipse (1 and 0.5);
\draw [-latex,ultra thick] (1,0.5) -- (1.1,0.5);
\draw [-latex,ultra thick] (1,1.5) -- (1.1,1.5);
\node[black, thick] at (1,-1) {$(a)$};
\end{tikzpicture}
\hspace{1.5cm}
\begin{tikzpicture}[scale=0.8]
\draw [fill](0,1) circle (0.2);
\draw [fill](2,1) circle (0.2);
\draw[style,ultra thick] (1,1) ellipse (1 and 0.5);
\draw [-latex,ultra thick] (0.8,0.5) -- (0.9,0.5);
\draw [-latex,ultra thick] (0.8,1.5) -- (0.9,1.5);
\draw [-latex,ultra thick] (1.2,0.5) -- (1.1,0.5);
\draw [-latex,ultra thick] (1.2,1.5) -- (1.1,1.5);
\draw [-, ultra thick] (0.85,1.35) -- (1.15,1.65);
\draw [-, ultra thick] (0.85,1.65) -- (1.15,1.35);
\draw [-, ultra thick] (0.85,0.35) -- (1.15,0.65);
\draw [-, ultra thick] (0.85,0.65) -- (1.15,0.35);
\node[black, thick] at (1,-1) {$(b)$};
\end{tikzpicture}
\hspace{1.5cm}
\begin{tikzpicture}[scale=0.8]
\draw [-, ultra thick] (-0.1,0.85) -- (0.2,1.15);
\draw [-, ultra thick] (-0.1,1.15) -- (0.2,0.85);
\draw [-, ultra thick] (-1,1.85) -- (-0.55,2.05);
\draw [-, ultra thick] (-0.8,2.2) -- (-0.8,1.7);
\draw [-, ultra thick] (1.1,1.85) -- (0.55,2.05);
\draw [-, ultra thick] (0.8,2.2) -- (0.8,1.7);
\draw [fill](0,3) circle (0.2);
\draw [fill](-1.5,1) circle (0.2);
\draw [fill](1.5,1) circle (0.2);
\draw [-latex,ultra thick] (-1,1) -- (-0.2,1);
\draw [-latex,ultra thick] (1,1)--(0.2,1) ;
\draw [-latex,ultra thick] (-1.5,1)--(-0.9,1.8);
\draw [-latex,ultra thick] (0,3)--(-0.66,2.1);
\draw [-latex,ultra thick] (0,3)--(0.66,2.1);
\draw [-latex,ultra thick] (1.5,1)--(0.9,1.8);
\draw [-, ultra thick] (-1.5,1) -- (0,3);
\draw [-, ultra thick] (1.5,1) -- (0,3);
\draw[-,ultra thick] (-1.5,1)--(1.5,1);
\node[black, thick] at (0,0) {$(c)$};
\end{tikzpicture}
\hspace{1.5cm}
\begin{tikzpicture}[scale=0.8]
\draw [-, ultra thick] (-0.15,0.85) -- (0.15,1.15);
\draw [-, ultra thick] (-0.15,1.15) -- (0.15,0.85);
\draw [-, ultra thick] (1.85,0.85) -- (2.15,1.15);
\draw [-, ultra thick] (1.85,1.15) -- (2.15,0.85);
\draw [-, ultra thick] (0.85,1.85) -- (1.15,2.15);
\draw [-, ultra thick] (0.85,2.15) -- (1.15,1.85);
\draw [-, ultra thick] (0.85,-0.15) -- (1.15,0.15);
\draw [-, ultra thick] (0.85,0.15) -- (1.15,-0.15);
\draw [fill](0,0) circle (0.2);
\draw [fill](0,2) circle (0.2);
\draw [fill](2,0) circle (0.2);
\draw [fill](2,2) circle (0.2);
\draw [-latex,ultra thick] (0,0) -- (0,0.8);
\draw [-latex,ultra thick] (0,2) -- (0,1.2);
\draw [-latex,ultra thick] (2,0) -- (2,0.8);
\draw [-latex,ultra thick] (2,2) -- (2,1.2);
\draw [-latex,ultra thick] (0,0) -- (0.8,0);
\draw [-latex,ultra thick] (2,0) -- (1.2,0);
\draw [-latex,ultra thick] (0,2) -- (0.8,2);
\draw [-latex,ultra thick] (2,2) -- (1.2,2);
\draw[-,ultra thick] (0,0)--(2,0);
\draw[-,ultra thick] (0,0)--(0,2);
\draw[-,ultra thick] (2,0)--(2,2);
\draw[-,ultra thick] (0,2)--(2,2);
\node[black, thick] at (1,-1) {$(d)$};
\end{tikzpicture}
\\
\begin{tikzpicture}[scale=0.8]
\draw [-, ultra thick] (1.8,0.85) -- (2.2,1.1);
\draw [-, ultra thick] (1.8,1.1) -- (2.2,0.85);
\draw [fill](0,1) circle (0.2);
\draw[style,ultra thick] (1,1) ellipse (1 and 0.5);
\draw [-latex,ultra thick] (0.9,0.5) -- (0.8,0.5);
\draw [-latex,ultra thick] (0.9,1.5) -- (0.8,1.5);
\node[black, thick] at (1,-1) {$(e)$};
\end{tikzpicture}
\hspace{1.5cm}
\begin{tikzpicture}[scale=0.8]
\draw [fill](0,1) circle (0.2);
\draw [fill](2,1) circle (0.2);
\draw[style,ultra thick] (1,1) ellipse (1 and 0.5);
\draw [-latex,ultra thick] (0.6,0.53) -- (0.5,0.54);
\draw [-latex,ultra thick] (0.6,1.49) -- (0.5,1.47);
\draw [-latex,ultra thick] (1.4,0.53) -- (1.5,0.54);
\draw [-latex,ultra thick] (1.4,1.49) -- (1.5,1.47);
\draw [-, ultra thick] (0.85,1.35) -- (1.15,1.65);
\draw [-, ultra thick] (0.85,1.65) -- (1.15,1.35);
\draw [-, ultra thick] (0.85,0.35) -- (1.15,0.65);
\draw [-, ultra thick] (0.85,0.65) -- (1.15,0.35);
\node[black, thick] at (1,-1) {$(f)$};
\end{tikzpicture}
\hspace{1.5cm}
\begin{tikzpicture}[scale=0.8]
\draw [-, ultra thick] (-0.1,0.85) -- (0.2,1.15);
\draw [-, ultra thick] (-0.1,1.15) -- (0.2,0.85);
\draw [-, ultra thick] (-1,1.85) -- (-0.55,2.05);
\draw [-, ultra thick] (-0.8,2.2) -- (-0.8,1.7);
\draw [-, ultra thick] (1.1,1.85) -- (0.55,2.05);
\draw [-, ultra thick] (0.8,2.2) -- (0.8,1.7);
\draw [fill](0,3) circle (0.2);
\draw [fill](-1.5,1) circle (0.2);
\draw [fill](1.5,1) circle (0.2);
\draw [-latex,ultra thick] (-1,1) -- (0.7,1);
\draw [-latex,ultra thick] (1,1)--(-0.6,1);
\draw [-latex,ultra thick] (-1.5,1)--(-0.36,2.5);
\draw [-latex,ultra thick] (0,3)--(-1.26,1.3);

\draw [-latex,ultra thick] (0,3)--(1.3,1.3);
\draw [-latex,ultra thick] (1.5,1)--(0.38,2.5);
\draw [-, ultra thick] (-1.5,1) -- (0,3);
\draw [-, ultra thick] (1.5,1) -- (0,3);
\draw[-,ultra thick] (-1.5,1)--(1.5,1);
\node[black, thick] at (0,0) {$(g)$};
\end{tikzpicture}
\hspace{1.5cm}
\begin{tikzpicture}[scale=0.8]
\draw [-, ultra thick] (-0.15,0.85) -- (0.15,1.15);
\draw [-, ultra thick] (-0.15,1.15) -- (0.15,0.85);
\draw [-, ultra thick] (1.85,0.85) -- (2.15,1.15);
\draw [-, ultra thick] (1.85,1.15) -- (2.15,0.85);
\draw [-, ultra thick] (0.85,1.85) -- (1.15,2.15);
\draw [-, ultra thick] (0.85,2.15) -- (1.15,1.85);
\draw [-, ultra thick] (0.85,-0.15) -- (1.15,0.15);
\draw [-, ultra thick] (0.85,0.15) -- (1.15,-0.15);
\draw [fill](0,0) circle (0.2);
\draw [fill](0,2) circle (0.2);
\draw [fill](2,0) circle (0.2);
\draw [fill](2,2) circle (0.2);
\draw [-latex,ultra thick] (0,0) -- (0,1.6);
\draw [-latex,ultra thick] (0,2) -- (0,0.4);
\draw [-latex,ultra thick] (2,0) -- (2,1.6);
\draw [-latex,ultra thick] (2,2) -- (2,0.4);
\draw [-latex,ultra thick] (0,0) -- (1.6,0);
\draw [-latex,ultra thick] (2,0) -- (0.4,0);
\draw [-latex,ultra thick] (0,2) -- (1.6,2);
\draw [-latex,ultra thick] (2,2) -- (0.4,2);
\draw[-,ultra thick] (0,0)--(2,0);
\draw[-,ultra thick] (0,0)--(0,2);
\draw[-,ultra thick] (2,0)--(2,2);
\draw[-,ultra thick] (0,2)--(2,2);
\node[black, thick] at (1,-1) {$(h)$};
\end{tikzpicture}
\caption{Feynman diagrams for Majoron mass, where the dot represent the vertex given in Eq.~(\ref{xxx}) and the cross represent the mass insertion. \label{feynmand}}
\end{center}
\end{figure}

We work in the framework of the SM extended by three right-handed sterile neutrinos $N_R^{} $ and one Higgs singlet $\Phi $ that carries two unit \textbf{L}. The general beyond SM Lagrangian can be written as
\begin{eqnarray}
\label{eq:Lagrangian}
{\cal L}_{\rm BSM}^{}  = \overline{N_R^{} } i \slashed{\partial } N_R^{} + \left(\partial_\mu \Phi \right)^\dagger (\partial^\mu \Phi) + \mu_\Phi^2 \Phi^\dagger \Phi^{}  - \lambda_1 (\Phi^\dagger \Phi)^2 -\lambda_2 (\Phi^\dagger \Phi) (H^\dagger H) \nonumber \\ -\left[  Y_{\rm N}^{}  \overline{\ell_L^{} } \tilde H N_R^{}  + {1\over 2}\overline{N_R^C} \left( Y_{\rm M}^{} \Phi + m  \right)N_R^{}  + {\rm h.c.}  \right]
\end{eqnarray}
where $\ell_L^{}$ is the left-handed lepton doublet, $H$ is the SM Higgs doublet, $Y_{\rm N, M}^{} $ is the Yukawa coupling matrix, $m$ is the LNV Majorana mass added to the Lagrangian considering that global symmetries in SM can be explicitly broken by quantum gravity effect. We define $\Phi ={1\over \sqrt{2} } \phi e^{i\theta}={1\over \sqrt{2} } \phi e^{ia/f_a}$ where $a$ is the Majoron with $f_a$ the decay constant, then the phase  in the Lagrangian can be removed by the field dependent phase transformation: $f_{L,R}^{} \to  e^{-i {\theta /2}}f_{L,R}^{}$ where $f$ represents leptons,  except that in the term $\overline{N_R^C} m N_R^{}  $. It has 
\begin{eqnarray}
{1\over 2} e^{-i \theta } \overline{N_R^C} m N_R^{} + {\rm h.c.} \; ,  \label{xxx}
\end{eqnarray}
which gives the new interaction of the Majoron. Radiative corrections induced by this interaction will give rise to an effective potential for the Majoron at the one-loop level, with the relevant Feynman diagrams given the Fig.~\ref{feynmand}.  Explicit calculation gives 
\begin{eqnarray}
\label{eq:potential}
V_a \sim  - {1\over 16\pi^2} \sum_{n=1}^4 a_n \cos {n \theta }. 
\end{eqnarray}
where $\theta \equiv a /f_a$,  $a_n$ are coefficients calculated from the Feynman diagrams in Fig.~\ref{feynmand} with their explicit expression listed in the Table \ref{tab:Couplings}.  We adopt the $\overline{\rm MS}$ renormalization scheme in the calculation.   Expanding the potential to the second order, one gets the mass of the Majoron
\begin{eqnarray}
m_a^2=\frac{1}{f_a^2}\frac{d^2V}{d\theta^2}= {1\over 16\pi^2f_a^2}\bigg|a_1+4a_2+9a_3+16a_4\bigg|.
\end{eqnarray}
Considering that $ m\ll M$ where $M$ is the mass of right-handed neutirno, $a_1$ will dominate the contribution to the mass of Majoron at low energy. This is a novel mechanism for the mass of goldstone bosons and it can be applied to the address the mass of ALPs in addition to the traditional misalignment mechanism~\cite{Dine:1982ah, Preskill:1982cy,Abbott:1982af,Co:2019jts,Chang:2019tvx,Batell:2021ofv,Chun:2021uwr,Chao:2022blc}. 

To derive the relic abundance of the Majoron, we need to solve the equation of motion (EOM) of the homogeneous Majoron field ($\theta \equiv a/ f_a$) in the FRW Universe \cite{Dine:1982ah, Preskill:1982cy, Abbott:1982af}, 
\begin{eqnarray}
\label{eq:EOM}
\ddot \theta + 3 H \dot \theta +\frac{1}{f_a^2} {d V_a \over d \theta } =0,
\end{eqnarray}
which is the same as the EOM of QCD axion except the initial condition, $\dot \theta_i \neq 0$. In our case,  the $\textbf{L}$ is explicitly broken by the Majorana mass term and $\dot \theta_i$ is nonzero, which is similar to the case of kinetic misalignment mechanism. The initial velocity can be obtained with the help of the Noether theorem~\cite{Co:2019wyp,Noether:1918zz}
\begin{eqnarray}
\label{eq:Noecharge}
\dot \theta^2 \propto {{\rm Tr} [Y_M^4 ]\over 96 \pi^2 } f_a^2 \cos 4\theta ,
\end{eqnarray}
in which we have ignored contributions of sub-leading terms. In the conventional misalignment mechanism, the Majoron is frozen at the initial value by the Hubble friction in the early Universe when $3 H > m_a$ and behaves as dark energy. It starts to oscillate and behaves as ordinary matter at the temperature $T^{\rm con}_{\rm osc}$ defined by $3 H (T^{\rm con}_{\rm osc})= m_a(T^{\rm con}_{\rm osc})$~\cite{Marsh:2015xka,DiLuzio:2020wdo,Co:2019jts}. 
%
However, a nonzero velocity $\dot \theta_i$ can delay the oscillation time given above \cite{Co:2019jts,Chang:2019tvx}, implying that $T^{\rm k}_{\rm osc}<T^{\rm con}_{\rm osc}$ with $T^{\rm k}_{\rm osc}$  the oscillation temperature in the case of kinetic misalignment mechanism. 
Notice that the terms containing $a_i (i=2,3,4)$ in the potential are of little importance compared to $a_1$ due to the extremely small parameter $m$. Thus, the maximum of the potential can be approximated by $V_{a,{\rm max}}\backsimeq  m_a^2 f_a^2$. Then  $T^{\rm k}_{\rm osc}$ can be roughly estimated by setting ${1}/{2}\dot \theta(T^{\rm k}_{\rm osc})^2f_a^2=V_{a,{\rm max}}$ \cite{Co:2019jts,Chang:2019tvx}, which gives
\begin{eqnarray}
\dot \theta (T^{\rm k}_{\rm osc})=\sqrt{2} m_a. 
\label{kine:tem}
\end{eqnarray}
Notice that there is a critical initial velocity, $\dot {\theta_i}_{\rm C}$, in which one obtains the same oscillation temperature from both scenarios, namely $T^{\rm k}_{\rm osc}=T^{\rm con}_{\rm osc}$. 

\begin{figure}[t]
\centering
\includegraphics[width=7.5cm]{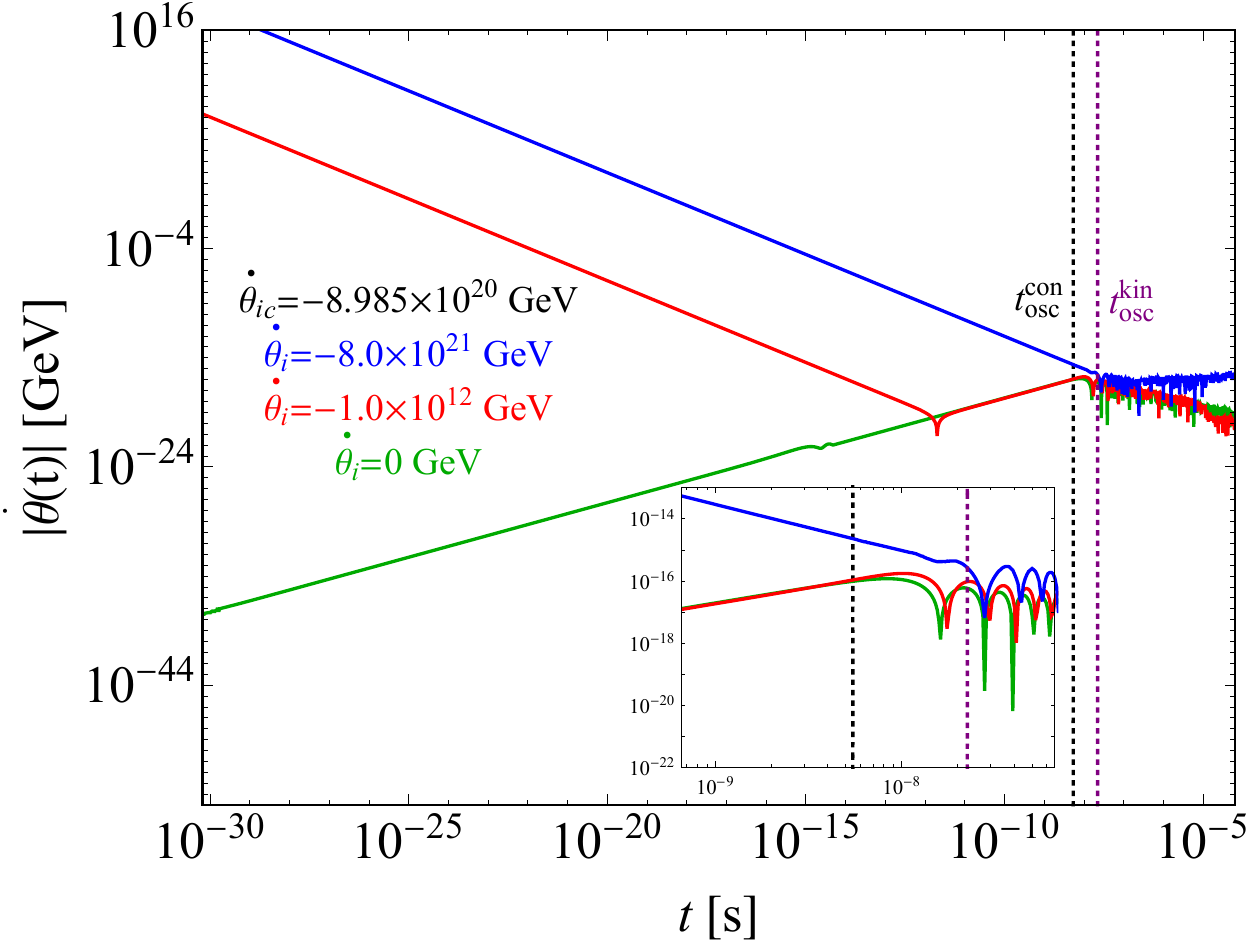}
\quad
\includegraphics[width=7.5cm]{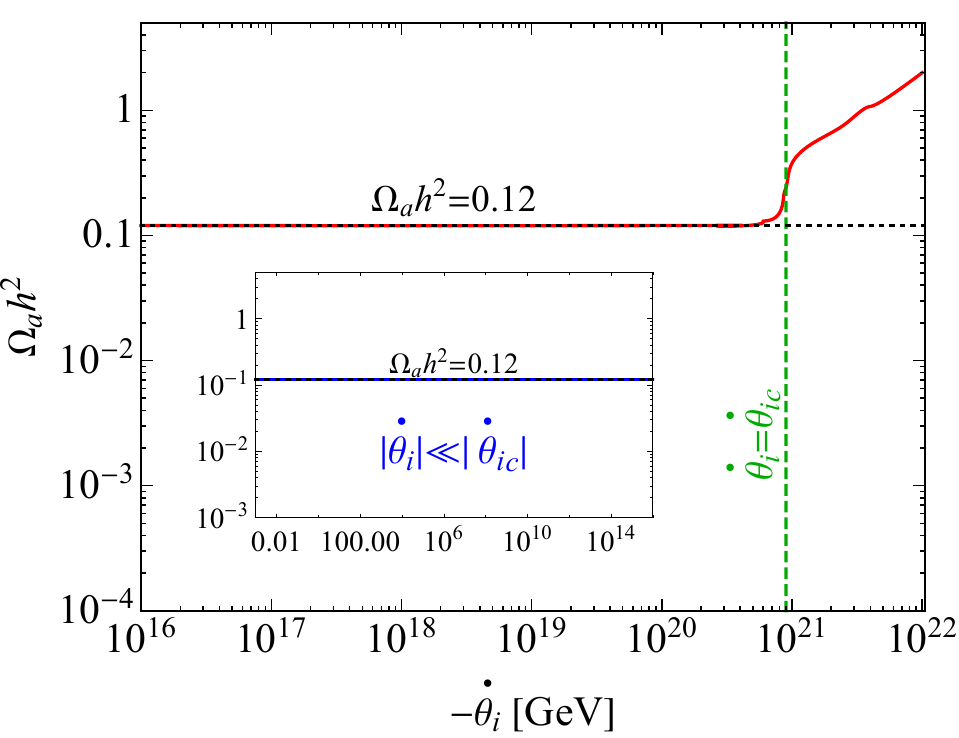}
\caption{Left panel: The evolution of Majoron velocity as a function of time. Three cases of $\dot \theta_i$ are depicted for comparison. The vertical dashed lines stand for the relevant oscillation times. The inset shows the evolution near the onset of oscillation. Right panel: The Majoron relic density as the function of initial velocity $\dot \theta_i$. The case for $\dot \theta_i\ll \dot {\theta_i}_{\rm C}$ is depicted in the inset. The vertical green dashed line represents the critical initial velocity $\dot {\theta_i}_{\rm C}$. The benchmark parameters are set to be $m=3\times 10^{-36}$ GeV, $M=10^{10}$ GeV, $f_a=10^{13}$ GeV.\label{fig:velrelic}}
\end{figure}

Solving the EOM in Eq.(\ref{eq:EOM}) numerically, we show the time evolutions of Majoron velocity with three benchmark initial values in the left panel of the Fig. \ref{fig:velrelic}. The conventional scenario in which $\dot \theta_i =0$ is depicted by the green curve for comparison. 
It shows that a large initial velocity ($\dot \theta_i >{\theta_i}_{\rm C}$) can not only significantly affect the velocity evolution, but also delay the oscillation time. Although a moderate initial velocity ($\dot \theta_i <{\theta_i}_{\rm C}$) can also alter the Majoron evolution in the early universe, it has little impact on the oscillation time  which is consistent with the case of $\dot \theta_i=0$.


Given the oscillation temperature, the energy density of the Majoron at present can be written  as
\begin{align}
\label{crit:temp}
\rho_a (T_0)
= \begin{cases}
\dfrac{1}{2} m_a^2 f_a^2 \langle  \theta_{a, i}^2 \rangle \dfrac{g_{*s} (T_0) }{g_{*s} (T^{\rm con}_{\rm osc})} \left( \dfrac{T_0 }{T^{\rm con}_{\rm osc}} \right)^3 \, ,\quad &\dot \theta_i\ll \dot {\theta_i}_{\rm C}\\
 m_a^2 f_a^2 \langle  \theta_{a, i}^2 \rangle \dfrac{g_{*s} (T_0) }{g_{*s} (T^{\rm k}_{\rm osc})} \left( \dfrac{T_0 }{T^{\rm k}_{\rm osc}} \right)^3\, ,\quad   &\dot \theta_i \geq{ \theta_i}_{\rm C}
\end{cases} 
\end{align}
where $g_{*s}$ is the relativistic degrees of the freedom, $T_0$ is the CMB temperature at present time and the angle brackets represent the value of the initial $\theta$ averaged over $[-\pi,~\pi)$~\cite{DiLuzio:2020wdo}. Here we consider the post-inflationary case and set the initial value to be $\theta(t_i) =\sqrt{\langle\theta_{a,i}^2\rangle}=\pi/\sqrt{3}$ \cite{Ringwald:2014vqa,DiLuzio:2020wdo}. The relic abundance of the DM is determined by normalizing the energy density to the critical energy density, i.e., $\Omega_a h^2 = \left[\rho_a(T_0)/\rho_{c,0}\right]h^2$ \cite{Marsh:2015xka,DiLuzio:2020wdo}. We show in the right panel of the Fig. \ref{fig:velrelic} the relic abundance of the Majoron as the function of $\dot \theta_i$, in which the horizontal dashed line represents the observed DM relic abundance $\Omega_a h^2\simeq0.12$~\cite{ParticleDataGroup:2022pth,Planck:2018vyg}. It shows that a sufficiently large initial velocity ($\dot \theta_i \gtrsim{\theta_i}_{\rm C}$) can significantly enhance the relic density of the Majoron, while medium or small initial velocities (in comparison to the critical initial velocity $\dot {\theta_i}_{\rm C}$) have little effect on $\Omega_a h^2$. In summary, the kinetic effect of Majoron only works when the initial velocity is close to or above the critical velocity ($\dot \theta_i \gtrsim{\theta_i}_{\rm C}$).

\section{Baryon asymmetry of the Universe}

The observed BAU is a longstanding problem to be addressed in cosmology. 
Here we shall show that a non-zero initial velocity $\dot \theta_i$ can not only significantly affect the evolution of Majoron field, but also  can be taken as an additional effective chemical potential biasing the BAU via the spontaneous baryogenesis mechanism, following the Refs.~\cite{Co:2020jtv,Domcke:2020kcp}, which provide a systematic study on the BAU arising from axions with generic couplings and a non-zero velocity. 
The Majoron enters the transport equations as a source term, 
\begin{eqnarray}
- \frac{d}{d \ln T} \left(  \frac{\mu_i}{T} \right)  = -\frac{1}{g_i}\sum_\alpha n_{i}^{\alpha} \frac{\gamma_\alpha}{H}
\left[\sum_{j}n_{j}^{\alpha} \left(  \frac{\mu_j}{T} \right)  - n_{S}^{\alpha}  \frac{\dot\theta(T)}{T} \right],
\label{eq:transport_eqs}
\end{eqnarray}
where $\mu_{i,j}$ is the chemical potential related to  particles of the kind $i$ and $j$, $g_i$ is the effective number of degrees of freedom, $n^{\alpha}_{i,j}$ is the vector indicating the charge of the particle $i$ for  the operator $O_{\alpha} $ (Here $O_{\alpha} $ are  Chern-Simons or Yukawa operators), $H$ is the Hubble parameter, and $\gamma_{\alpha}\equiv{6\Gamma_{\alpha}}/{T^{3}}$ with $\Gamma_{\alpha}$ the transport rate per unit time-volume. It should be mentioned that the source vector $n_{S}^{\alpha}$ is determined by the specific interactions of the Majoron.

To calculate the BAU induced by the Majoron,  we first need to write down the explicit form of charge vectors as well as source vectors.  The particle species $i$'s can be taken as~\cite{Domcke:2020kcp}
\begin{eqnarray}
\label{eq:SMspecies}
i = \tau, ~L_{12}, ~L_3, ~q_{12}, ~t, ~b, ~Q_{12}, ~Q_3, ~H,
\end{eqnarray}
with effective degrees of freedom given as $g_i =\left( 1, 4, 2, 12, 3, 3, 12, 6, 4\right)$. As mentioned in the introduction, the small neutrino masses  arise from the dimension-5 Weinberg operator~\cite{Weinberg:1979sa}, $-\frac{1}{2M_5} \ell \ell H H$, which is known arising from the decouple of the seesaw particles~\cite{Yanagida:1979as,GellMann:1979vob,Minkowski:1977sc,Mohapatra:1979ia}. 
\begin{figure}[t]
\begin{center}
\begin{tikzpicture}[scale=0.8]
\draw [-, ultra thick] (0.85,-0.15) -- (1.15,0.15);
\draw [-, ultra thick] (0.85,0.15) -- (1.15,-0.15);
\draw [-, ultra thick] (2.85,-0.15) -- (3.15,0.15);
\draw [-, ultra thick] (2.85,0.15) -- (3.15,-0.15);
\draw [-latex, ultra thick] (-1,0) -- (-0.50,0);
\draw [-latex, ultra thick] (5,0) -- (4.5,0);
\draw [-latex,ultra thick] (0,0) -- (0.8,0);
\draw [-latex,ultra thick] (2,0) -- (1.2,0);
\draw [-latex,ultra thick] (2,0) -- (2.8,0);
\draw [-latex,ultra thick] (4,0) -- (3.2,0);
\draw[-,ultra thick] (0,0)--(2,0);
\draw[-,ultra thick,dotted] (0,0)--(0,2);
\draw[-,ultra thick,dashed] (2,0)--(2,2);
\draw[-,ultra thick] (2,0)--(4,0);
\draw[-,ultra thick] (-1.5,0)--(0,0);
\draw[-,ultra thick] (4,0)--(5.5,0);
\draw[-,ultra thick,dotted] (4,0)--(4,2);
\node[red, thick] at (-1,0.5) {$\ell_L$};
\node[red, thick] at (5,0.5) {$\ell_L$};
\node[red, thick] at (1,0.5) {$N_R$};
\node[red, thick] at (3,0.5) {$N_R$};
\node[red, thick] at (2,2.2) {$\Phi$};
\node[red, thick] at (0,2.2) {$H$};
\node[red, thick] at (4,2.2) {$H$};
\end{tikzpicture}
\caption{Feynman diagram for the Majoron-Weinberg operator interaction. \label{fig:MajorWeinberg}}
\end{center}
\end{figure}
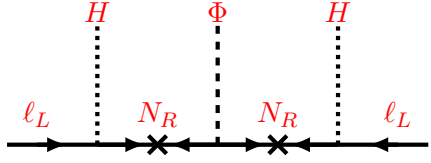
Based on the Lagrangian given in Eq. \eqref{eq:Lagrangian}, the Feynman diagram shown in Fig. \ref{fig:MajorWeinberg} induces a Majoron-Weinberg operator interaction  of the form
\begin{eqnarray}
\mathcal{L}_{\rm int}\supset\frac{1}{2M} \frac{a}{f_a}\ell \ell H H,
\end{eqnarray}
which provides an additional source term  for the transport equations.
Although this interaction is highly suppressed by the heavy neutrino masses, it is  in thermal equilibrium in the very early time due to the extremely high temperature of the universe. 
Once the transport rate per unit time for the Weinberg operator becomes lower than the Hubble parameter, $\sum_i \frac{1}{g_i} \left(  n_{i}^W \right) ^2  \gamma_W < H$, it decouples from the thermal bath. The critical condition (5$\gamma_{W}= H$) gives the decoupling temperature as~\cite{Domcke:2020kcp,Co:2020jtv}
\begin{eqnarray}
T_\mathrm{W} \simeq 6\times 10^{12}\,\mathrm{GeV}\times \left(\frac{0.05\,\mathrm{eV}}{m_\nu}\right)^2. 
\label{eq:T_W}
\end{eqnarray}
Obviously,  the Weinberg operator plays an important rule in the transport equations only for the temperature above $T_W$. 

As shown in the Sec. \ref{sec:MajoronDM},  interactions of  Majoron with leptons and electroweak gauge bosons can be obtained from the following field dependent phase transformations: $f_{L,R}^{} \to  e^{-i {a /2f_a}}f_{L,R}^{}$  and $\Phi \to  e^{i {a/f_a }} \Phi $, which results in
\begin{eqnarray}
\label{eqapp:action}
S\supset\int d^4x\left[\mathcal{L}_{\rm kin}-\frac{a}{2f_a}\partial_\mu(\overline{\ell_L}\gamma^\mu\ell_L+\overline{E_R}\gamma^\mu E_R)\right],
\end{eqnarray}
where ${\cal L}_{\rm kin}$ is the conventional kinetic Lagrangian for leptons and the second term shows that the Majoron interacts with SM leptons through the derivative form of the lepton current, i.e, $\mathcal{L}_{\rm int} = -\frac{1}{2}\frac{a}{f_a}\partial_\mu J_L^\mu$ (Here we define $J_L^\mu=\overline{\ell_L}\gamma^\mu\ell_L+\overline{E_R}\gamma^\mu E_R$ for convenience). Hence the source vector is of the following form~\cite{Domcke:2020kcp}
\begin{eqnarray}
\label{eq:lepsource}
n_S^\alpha 
= \frac{1}{2}\sum_i n_i^{\textbf{L}} n_i^\alpha = \frac{1}{2}\left(n_{\tau}^{\alpha} +n_{L_{12}}^{\alpha} + n_{L_3}^{\alpha}\right),
\end{eqnarray}
where $n_i^\textbf{L} = \left(1, 1, 1, 0, 0, 0, 0, 0, 0\right) $ being the vector corresponding to the \textbf{L}. Combining Eq. (\ref{eq:lepsource}) with the charge vectors $n^{\alpha}_i$, the source vector is finally given by
\begin{eqnarray}
\label{eq:sourcevec}
\left( n_S^{WS}, n_S^{W_{12}}, n_S^{W_3}, n_S^{SS},n_S^{Y_\tau},n_S^{Y_t},n_S^{Y_b}\right) = \left(\frac{3}{2}, 1, 1,0,0,0,0\right).
\end{eqnarray}

\begin{figure}[h]
\centering
\includegraphics[width=7.9cm]{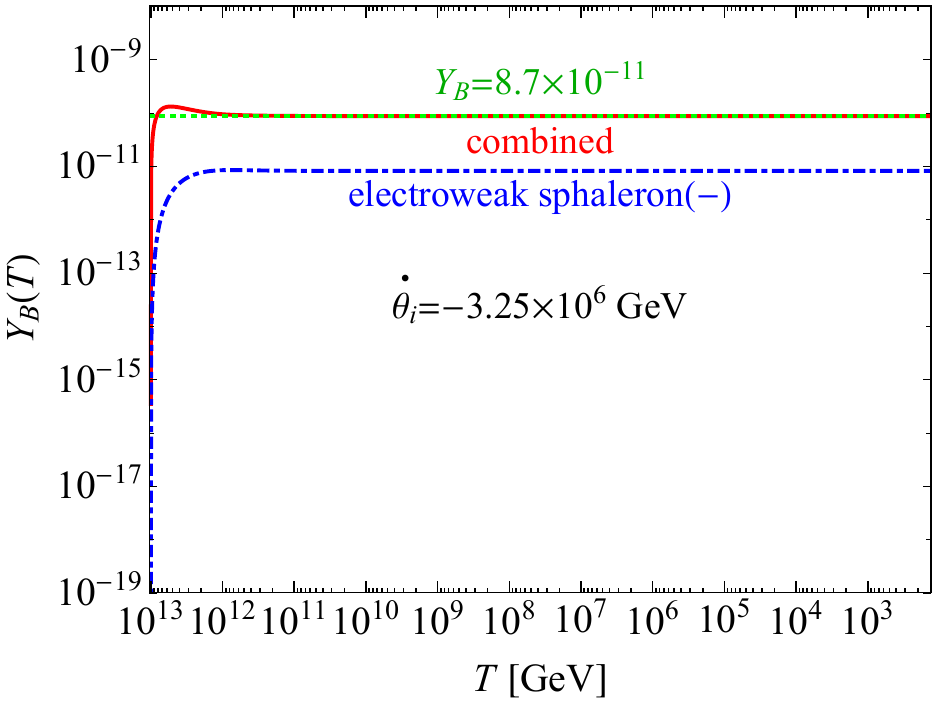}
\quad
\includegraphics[width=7.9cm]{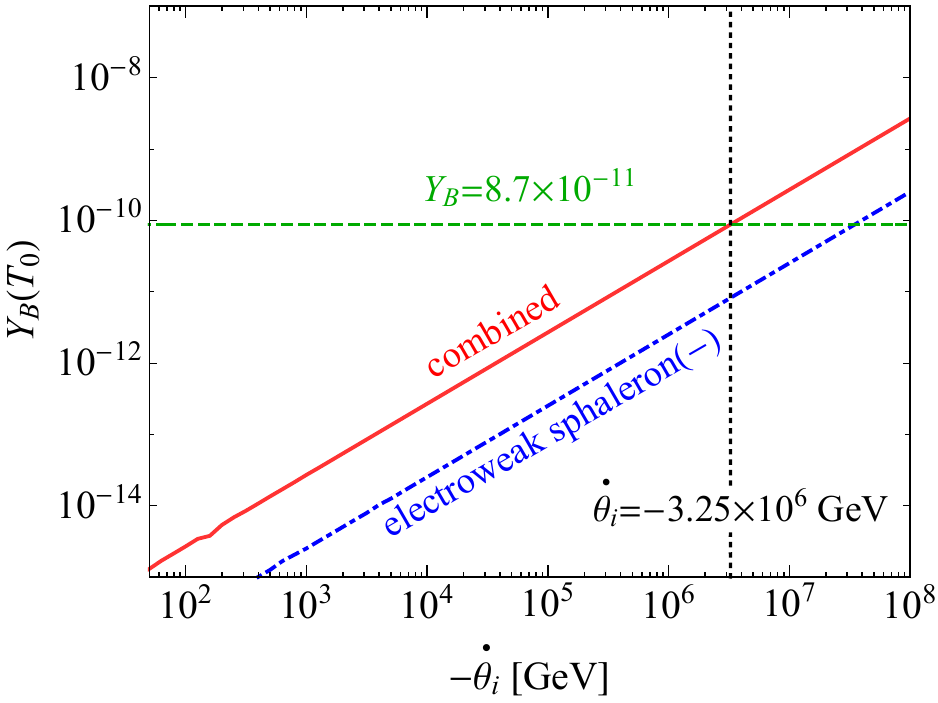}
\caption{Left panel: The baryon asymmetry evolves as a function of temperature. Right panel: The dependence of the baryon asymmetry today on initial velocity of Majoron. The dashed green line represents the observed baryon asymmetry $Y_B=8.7\times10^{-11}$\cite{Planck:2018vyg}.\label{fig:bau} The red and blue lines indicate the combined baryon asymmetry and the absolute value of baryon asymmetry only produced by the electroweak sphaleron process, respectively. }
\end{figure}

Based on the discussion above, the explicit expression of the transport equations shown in Eq.~\eqref{eq:transport_eqs} are obtained to be
\begin{align}
-\frac{d}{d\ln T}\left( \frac{\mu_\tau}{T}\right)=&-\frac{1}{g_\tau}(-1)\frac{\gamma_{Y_\tau}}{H}\left(-\frac{\mu_\tau}{T}+\frac{\mu_{L_3}}{T}+\frac{\mu_H}{T}\right) ,\nonumber\\
-\frac{d}{d\ln T}\left( \frac{\mu_{L_{12}}}{T}\right)=&-\frac{1}{g_{L_{12}}}2\frac{\gamma_{WS}}{H}\left[\frac{2\mu_{L_{12}}}{T}+\frac{\mu_{L_3}}{T}+\frac{6\mu_{Q_{12}}}{T}+\frac{3\mu_{Q_3}}{T}-\frac{3}{2}\frac{\dot\theta(T)}{T} \right]\nonumber\\
&-\frac{1}{g_{L_{12}}}\frac{2\gamma_{W_{12}}}{H}\left[\frac{2\mu_{L_{12}}}{T}+\frac{2\mu_H}{T}-\frac{\dot\theta(T)}{T} \right],\nonumber\\
-\frac{d}{d\ln T}\left( \frac{\mu_{L_3}}{T}\right)=&-\frac{1}{g_{L_3}}\frac{\gamma_{WS}}{H}\left[\frac{2\mu_{L_{12}}}{T}+\frac{\mu_{L_3}}{T}+\frac{6\mu_{Q_{12}}}{T}+\frac{3\mu_{Q_3}}{T}-\frac{3}{2}\frac{\dot\theta(T)}{T} \right]\nonumber\\
~&-\frac{1}{g_{L_3}}\frac{\gamma_{{Y_\tau}}}{H}\left[-\frac{\mu_\tau}{T}+\frac{\mu_{L_3}}{T}+\frac{\mu_H}{T}\right]-\frac{1}{g_{L_3}}\frac{2\gamma_{W_3}}{H}\left[\frac{2\mu_{L_3}}{T}+\frac{2\mu_H}{T}-\frac{\dot\theta(T)}{T} \right],\nonumber\\
-\frac{d}{d\ln T}\left( \frac{\mu_{q_{12}}}{T}\right)=&\frac{1}{g_{q_{12}}}\frac{4\gamma_{SS}}{H}\left[-\frac{4\mu_{q_{12}}}{T}-\frac{\mu_t}{T}-\frac{\mu_b}{T}+\frac{4\mu_{Q_{12}}}{T}+ \frac{2\mu_{Q_3}}{T} \right],\nonumber\\
-\frac{d}{d\ln T}\left( \frac{\mu_t}{T}\right)=&\frac{1}{g_t}\frac{\gamma_{SS}}{H}\left[-\frac{4\mu_{q_{12}}}{T}-\frac{\mu_t}{T}-\frac{\mu_b}{T}+\frac{4\mu_{Q_{12}}}{T}+ \frac{2\mu_{Q_3}}{T}\right]\nonumber\\
~&+\frac{1}{g_t}\frac{\gamma_{Y_t}}{H}\left[-\frac{\mu_t}{T}+\frac{\mu_{Q_3}}{T}+\frac{\mu_H}{T} \right],\nonumber\\
-\frac{d}{d\ln T}\left( \frac{\mu_b}{T}\right)=&\frac{1}{g_b}\frac{4\gamma_{SS}}{H}\left[-\frac{4\mu_{q_{12}}}{T}-\frac{\mu_t}{T}-\frac{\mu_b}{T}+\frac{4\mu_{Q_{12}}}{T}+ \frac{2\mu_{Q_3}}{T}\right]\nonumber\\
~&+\frac{1}{g_b}\frac{\gamma_{Y_b}}{H}\left[-\frac{\mu_b}{T}+\frac{\mu_{Q_3}}{T}-\frac{\mu_H}{T}\right],\nonumber\\
-\frac{d}{d\ln T}\left( \frac{\mu_{Q_{12}}}{T}\right)=&-\frac{1}{g_{Q_{12}}}\frac{4\gamma_{SS}}{H}\left[-\frac{4\mu_{q_{12}}}{T}-\frac{\mu_t}{T}-\frac{\mu_b}{T}+\frac{4\mu_{Q_{12}}}{T}+ \frac{2\mu_{Q_3}}{T}\right]
\nonumber\\
~&-\frac{1}{g_{Q_{12}}}\frac{6\gamma_{WS}}{H}\left[\frac{2\mu_{L_{12}}}{T}+\frac{\mu_{L_3}}{T}+\frac{6\mu_{Q_{12}}}{T}+\frac{3\mu_{Q_3}}{T}-\frac{3}{2}\frac{\dot\theta(T)}{T} \right],\nonumber\\
-\frac{d}{d\ln T}\left( \frac{\mu_{Q_3}}{T}\right)=&-\frac{1}{g_{Q_3}}\frac{2\gamma_{SS}}{H}\left[-\frac{4\mu_{q_{12}}}{T}-\frac{\mu_t}{T}-\frac{\mu_b}{T}+\frac{4\mu_{Q_{12}}}{T}+ \frac{2\mu_{Q_3}}{T}\right]
\nonumber\\
~&-\frac{1}{g_{Q_3}}\frac{3\gamma_{WS}}{H}\left[\frac{2\mu_{L_{12}}}{T}+\frac{\mu_{L_3}}{T}+\frac{6\mu_{Q_{12}}}{T}+\frac{3\mu_{Q_3}}{T}-\frac{3}{2}\frac{\dot\theta(T)}{T} \right]\nonumber\\
~&-\frac{1}{g_{Q_3}}\frac{\gamma_{Y_t}}{H}\left[-\frac{\mu_t}{T}+\frac{\mu_{Q_3}}{T}+\frac{\mu_{Q_H}}{T}\right]-\frac{1}{g_{Q_3}}\frac{\gamma_{Y_b}}{H}\left[-\frac{\mu_b}{T}+\frac{\mu_{Q_3}}{T}-\frac{\mu_{Q_H}}{T}\right],\nonumber\\
-\frac{d}{d\ln T}\left( \frac{\mu_H}{T}\right)=&-\frac{1}{g_H}\frac{\gamma_{Y_\tau}}{H}\left[-\frac{\mu_\tau}{T}+\frac{\mu_{L_3}}{T}+\frac{\mu_H}{T} \right]-\frac{1}{g_H}\frac{\gamma_{Y_t}}{H}\left[-\frac{\mu_t}{T}+\frac{\mu_{Q_3}}{T}+\frac{\mu_{Q_H}}{T}\right]
\nonumber\\
~&+\frac{1}{g_H}\frac{\gamma_{Y_b}}{H}\left[-\frac{\mu_b}{T}
+\frac{\mu_{Q_3}}{T}-\frac{\mu_{Q_H}}{T} \right]
-\frac{1}{g_H}\frac{2\gamma_{W_{12}}}{H}\left[\frac{2\mu_{L_{12}}}{T}+\frac{2\mu_H}{T}-\frac{\dot\theta(T)}{T} \right]
\nonumber\\
~&-\frac{1}{g_H}\frac{2\gamma_{W_3}}{H}\left[\frac{2\mu_{L_3}}{T}+\frac{2\mu_H}{T}-\frac{\dot\theta(T)}{T} \right],
\end{align}
where the transport rate per unit time $\gamma_{\alpha}\equiv{6\Gamma_{\alpha}}/{T^{3}}$ are given in the Table 1 of Ref.~\cite{Domcke:2020kcp}. 
These equations describe the transport of the  Majoron source term into the \textbf{B}$-$\textbf{L} asymmetry, which is subsequently transported into the BAU via the sphaleron process. To calculate the BAU, we first need to derive the \textbf{B}$-$\textbf{L} asymmetry~\cite{Domcke:2020kcp,Co:2020jtv}, which is defined as $n_{{\footnotesize \textbf{B}-\textbf{L}}}=\frac{T^2}{6} \mu_{\footnotesize \textbf{B}-\textbf{L}}$ with~\cite{Domcke:2020kcp}
\begin{eqnarray}
\mu_{{\footnotesize \textbf{B}-\textbf{L}}}=4\mu_{q_{12}}+\mu_t+\mu_b+4\mu_{Q_{12}}+2\mu_{Q_3}-(\mu_\tau+4\mu_{L_{12}}+2\mu_{L_3})\; ,
\end{eqnarray}
and $\mu_{{\footnotesize \textbf{B}-\textbf{L}}}$ is closely related to the BAU through~\cite{Kolb:1990vq,Harvey:1990qw,Xing:2011zza}
\begin{eqnarray}
 Y_{\footnotesize \textbf{B}} = \frac{n_{\footnotesize \textbf{B}}}{s} =  \frac{T^3}{6s} \frac{\mu_{\footnotesize \textbf{B}}}{T} =  \frac{T^3}{6s} \left( \frac{28}{79}\frac{\mu_{\footnotesize \textbf{B}-\textbf{L}}}{T}\right)  \simeq 0.001  \frac{\mu_{\footnotesize \textbf{B}-\textbf{L}}}{T}.
 \label{eq:B-LtoB}
\end{eqnarray}
For numerical calculations, we take the initial condition of the chemical potential for all particle species as $\mu_i (f_a)=0$, various couplings and the neutrino mass are set as~\cite{Staub:2013tta,Goertz:2023pvn},
$
g_2 = 0.55\,,
g_3 = 0.59\,,
y_\tau =0.01\,,
y_t = 0.50\,,
y_b = 8.1\times 10^{-3}\,,
m_\nu = 0.05\,\mathrm{eV}\,.
$
By solving the transport equations numerically, we show in Fig.~\ref{fig:bau} the evolution of the BAU as the function of the temperature and the initial velocity, respectively. Here the input parameters related to the Majoron  are consistent with those in the Fig. \ref{fig:velrelic}.

As shown in the left panel of the Fig.~\ref{fig:bau}, the BAU increases first and then decreases as the electroweak sphaleron process produces a negative BAU, which washes out a portion of the BAU produced by the Weinberg operator. Nevertheless, the BAU is conserved soon after the Weinberg operator is decoupled, indicating that the Weinberg operator is absolutely  the dominate process. This is due to the fact that the Weinberg operator solely breaks the \textbf{L} symmetry, the electroweak sphaleron is suppressed compared to it even though both of them have non-zero source terms~\cite{Domcke:2020kcp}. In the right panel of the Fig.~\ref{fig:bau}, we show the BAU at present as a function of the initial velocity. There is no doubt that a larger initial velocity results in a larger BAU, which is what a rough equilibrium solution tells us~\cite{Domcke:2020kcp}. Moreover, the observed BAU can be produced for $\dot\theta_i=-3.25\times 10^6 $ GeV, which is in good agreement with the result of temperature evolution in the left panel.

\section{Summary}
\label{sec:summ}
ALPs are  a class of promising DM candidates. However, their mass generation mechanism, relic abundance and impacts to other new physics still elude us. In this letter, we propose a novel mass generation mechanism for a special ALP, the Majoron, via radiative corrections induced by the Majorana mass term of right-handed neutrinos in the canonical seesaw mechanism, which also induce an initial velocity for the Majoron according to the Noether theorem. We studied the impact of the initial velocity of the Majoron to its relic density, which shows that only a relatively large initial velocity can cause effects on it. In the meanwhile we show that the same initial velocity may lead to the BAU via the spontaneous baryogenesis mechanism with the help of the Weinberg operator, which naturally exists in this scenario. This study provides a generic  mechanism for the mass generation of a class of ALPs with global U(1) symmetry in addition to the traditional misalignment mechanism, which may be extended to model buildings of ALPs induced by the well-known $U(1)_R$, $U(1)_B$ and $U(1)_{B\pm L}$ global symmetries, that may also address the  BAU. Furthermore, we would like to make a brief comment on the constraint or signal of the Majoron DM. The interaction of the form  $\frac{a}{2f_a}\partial_\mu\left( \overline{\ell_L}\gamma^\mu\ell_L+\overline{E_R}\gamma^\mu E_R\right) $ generates chiral anomalies in the SM lepton sector~\cite{Bell:1969ts,Adler:1969gk,Riotto:1999yt}, which show that the Majoron couples  with both neutrinos and electroweak gauge bosons, but not electrons and diphotons \cite{Chao:2022blc}. Nevertheless, the coupling between the Majoron and neutrinos is double suppressed by the tiny neutrino mass $m_\nu$ and the high $\mathbf{L}$ breaking scale $f_a$, which makes the matter effect of neutrino oscillations produced by  the Majoron insignificant and has little impact on the effective number of neutrinos $N_{\rm eff}$ \cite{Li:2023kuz}. However, the chiral anomalies induce a Majoron-$Z$-photon ($aZ\gamma$) interaction, leaving the inverse Primakoff process $a+{\rm Z}e\to\gamma+{\rm Z}e$~\cite{Paschos:1993yf,Buchmuller:1989rb,Creswick:1997pg} mediated by the $Z$ boson as a viable detection method to detect this type of Majoron signal in terrestrial experiments, which will be studied in a future work.

\begin{acknowledgments}
This work was supported by the National Natural Science Foundation of China (NSFC) (Grants No. 11775025 and No. 12175027). and the Fundamental Research Funds for the Central Universities under grant No. 2017NT17.
\end{acknowledgments}

\bibliography{references}

\end{document}